\documentclass[sigconf,nonacm]{acmart}

\AtBeginDocument{%
  \providecommand\BibTeX{{%
    \normalfont B\kern-0.5em{\scshape i\kern-0.25em b}\kern-0.8em\TeX}}}

\usepackage{multirow}
\usepackage{caption}
\usepackage{subcaption}
\newcommand{\R}{\mathbb{R}}
\usepackage{footnote}
\usepackage[
  font = small,
  labelfont = bf,
  tableposition = top
]{caption}
\newcommand{\etal}{\textit{et al}. }

\begin{document}

\title{Learning Similarity Preserving Binary Codes for \\ Recommender Systems}

\author{Yang Shi}
\email{yang.shi@rakuten.com}
\affiliation{%
  \institution{Rakuten Group,  Inc.}
  \city{San Mateo}
  \state{CA}
  \country{USA}
}

\author{Young-joo Chung}
\email{youngjoo.chung@rakuten.com}
\affiliation{%
 \institution{Rakuten Group, Inc.}
  \city{San Mateo}
  \state{CA}
  \country{USA}}

\renewcommand{\shortauthors}{Shi and Chung}

\begin{abstract}
Hashing-based Recommender Systems (RSs) are widely studied to provide scalable services. The existing methods for the systems combine three modules to achieve efficiency: feature extraction, interaction modeling, and binarization. In this paper, we study an unexplored module combination for the hashing-based recommender systems, namely Compact Cross-Similarity Recommender (CCSR). Inspired by cross-modal retrieval, CCSR utilizes Maximum a Posteriori similarity instead of matrix factorization and rating reconstruction to model interactions between users and items. We conducted experiments on MovieLens1M, Amazon product review, Ichiba purchase dataset and confirmed CCSR outperformed the existing matrix factorization-based methods. On the Movielens1M dataset, the absolute performance improvements are up to 15.69\% in NDCG and  4.29\% in Recall. In addition, we extensively studied three binarization modules: $sign$, scaled $tanh$, and sign-scaled $tanh$.  The result demonstrated that although differentiable scaled $tanh$ is popular in recent discrete feature learning literature, a huge performance drop occurs when outputs of scaled $tanh$ are forced to be binary.
\end{abstract}



\maketitle

\section{Introduction}
\label{sec:intro}
Neural recommender systems have dramatically improved the recommendation performance 
in recent years. However, it is difficult to scale them up due to their high computational cost~\cite{Lian2020:lightrec}. To overcome this issue,
hashing-based recommender systems have been widely studied. The binarized user and item representations not only reduce the memory requirement but also accelerate the recommendation speed. For example, 
if items are represented by 256-dim double-precision float vectors, 10 million items’ representations will take over 19 GB of storage space. With Hashing-based recommendation, 
it will only need 0.3 GB.

\begin{figure}[t]
     \centering
     \includegraphics[height=4.5cm]{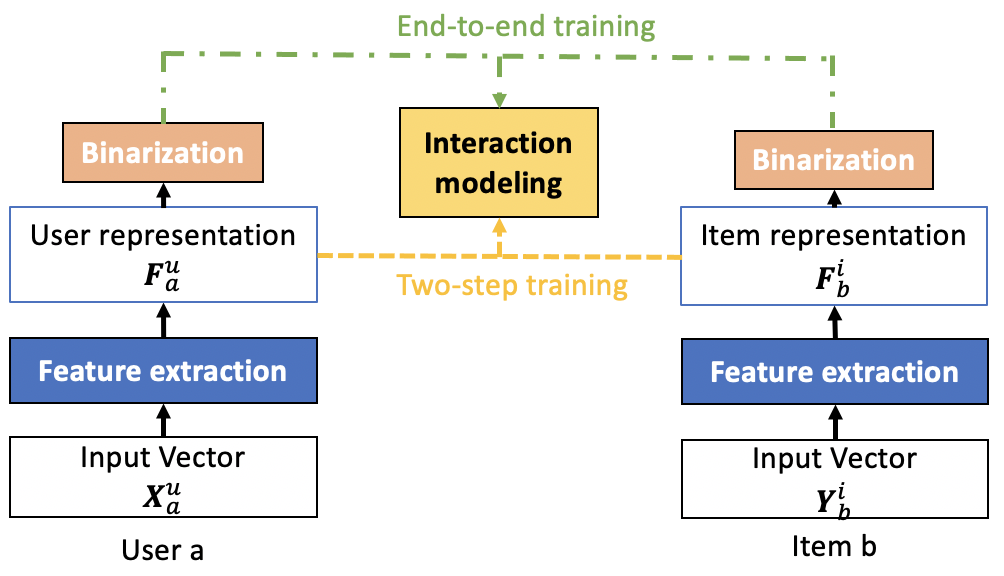}
     
     \caption{Hashing-based recommendation systems. Prediction is made by computing Hamming distances between binary codes.}
     \label{fig:model}
  \end{figure}

The existing methods for the hashing-based RSs can be divided into three modules: feature extraction, interaction modeling, and binarization. The feature extraction module takes user's implicit or explicit feedback on items as inputs and learns representations by modeling their interactions (i.e., user/item preferences). After learning these real-valued representations, binarization methods are applied to convert them into binary codes. These binarizations can be done after learning real-valued representation~\cite{bccf,PPH} (the two-step approach) or jointly done with feature learning~\cite{cccf,dpr} (the direct approach). The final recommendation will be conducted by measuring the distances of the codes in Hamming space. Figure~\ref{fig:model} shows the general framework for the hashing-based recommender system. 

  \begin{table}[t]
\centering
\caption{Comparison of different hashing-based recommender systems in terms of feature extraction, interaction modeling, and binarization}
    \scalebox{0.73}{
    \begin{tabular}[b]{|c|c|c|c|c|c|c|c|}
    \hline
        \multirow{3}{*}{Paper}  & \multicolumn{6}{c|}{Loss function} & \multirow{3}{*}{\textbf{Binarization}} \\
    \cline{2-7}
    &\multicolumn{3}{c|}{\textbf{Feature extraction}}&\multicolumn{3}{c|}{\textbf{User-item interaction}}&\\
    \cline{2-7}
    & MF & AE & Other NN  & Dot product & CE& MAP &\\
    \hline
    CFCodeReg\cite{bccf} &   \checkmark& & & \checkmark &&& $Sign$ \\
    \hline
    DCF\cite{discrete}  & \checkmark& && \checkmark& && $Sign$\\
    \hline
    NBR\cite{nbr}  & \checkmark&\checkmark && \checkmark& && LPR \\
        \hline
    NeuHash\cite{neuhash}  &  \checkmark & \checkmark &&\checkmark&&&STE \\
    \hline
    HashGNN\cite{stenew}  &&&\checkmark&&\checkmark& & $Sign$, STE \\
    \hline
    CCSR (ours)   & & \checkmark &  & & &\checkmark & $Sign$, Scaled $tanh$\\
    \hline
    \end{tabular}
    }
    \label{tab:modelcompare}
    \end{table}
  
With the success of Collaborative Filtering (CF) that creates the user/item representations based on their interactions, earlier works focused on combining hashing and CF~\cite{pmlr-v9-karatzoglou10a,bccf}. These approaches obtain real-valued user/item features using Matrix Factorization (MF) and then convert them into binary codes using Linear Programming Relaxation (LPR). Users’ preferences on items are modeled with dot-product between user/item representations.


Recently, several methods have been proposed to integrate neural networks and hashing for recommender systems~\cite{beyond,fashiontanh}. These methods utilize neural networks (e.g., autoencoders, graph neural networks, and convolutional neural networks) to obtain real-valued user/item representations and binarize them by a hard threshold ($sign$) operator or an approximation (scaled $tanh$) function or straight-through-estimator (STE). Users’ preferences on items are usually modeled with dot-product similarity, cross-entropy (CE) loss or rank loss. 

Table~\ref{tab:modelcompare} shows that previous works have different design choices on each module and certain combinations have not been explored. In this paper, we study an unexplored combination; we learn real-valued user/item representation through autoencoders and MAP-based similarity and obtain binary codes for these user/item representations. We call our method \textbf{C}ompact \textbf{C}ross-\textbf{S}imilarity \textbf{R}ecommender (CCSR).
Modeling entities with MAP-based similarity has shown its effectiveness in cross-modal retrieval (CR), where the goal of the system is retrieving similar entities from different modalities (e.g., image/text, audio/text).  The recommendation task can be considered as cross-modal retrieval where the user space is one modality and the item space is the other. From this viewpoint, user/item representations are created by mapping them to the shared latent space while preserving the original similarities (i.e., preference). In the recommendation time, we retrieve the k-most similar items to users by comparing user/item similarity in this shared space.

We further empirically study the role of different binarization methods in the final performance. To the best of our knowledge, the impact of changing binarization methods while fixing other modules has not been studied in the hashing-based recommender system literature. Previous works reported the results only based on a single binarization method.

Our contributions are as follows:
\begin{itemize}
    \item We categorize the previous hash-based RS methods based on their design choices and explore a new design, CCSR, which is inspired by cross-modal retrieval literature.
    \item We conclude that MAP-based similarity loss is better than MF-based rating reconstruction loss in the top-k recommendation task. This is because the former loss emphasizes more on similarity learning which is important for the recommendation task. The latter focuses on rating reconstruction which is an indirect approach to recommendation.
    \item CCSR model outperforms hashing-based  models with Matrix-factorization by up to  15.69\%, 1.53\% and 1.69\% NDCG absolute improvements on Movielens1M, Amazon and Ichiba datasets respectively.
    \item We show that different models prefer different binarization methods. The simple $sign$ function still performs well compared to other more complicated methods.
    Even though differentiable scaled $tanh$ is popular in recent discrete feature learning literature, a huge performance drop occurred when scaled $tanh$ outputs are forced to be binary. 
\end{itemize}
\section{Related Work}
\textbf{Hashing-based CF} Collaborative filtering utilizes observed user-item interaction (e.g., ratings, clicks, and purchases) to estimate unobserved interactions. 
Earlier works on hashing-based CF utilized MF and a two-stage binarization~\cite{bccf,PPH}. They first learned real-valued representations through MF and then converted them to binary codes. 
Several works proposed to use different loss functions, better binarization algorithms, or additional features for performance improvement. 

Compositional Coding for Collaborative Filtering (CCCF)\cite{cccf} and Discrete Collaborative Filtering (DCF) \cite{discrete} proposed learning binary codes directly using discrete bit-by-bit optimization. In addition to utilizing the discrete optimization, Discrete Personalized Ranking (DPR) \cite{dpr} changed the rating reconstruction loss to AUC-aimed loss; Discrete Content-aware Matrix Factorization (DCMF) \cite{dcmf} added a content-specific feature with the learned item feature. Hansen \etal~\cite{10.1145/3442381.3450011} proposed a new method that can project the dissimilarity between two objects into Hamming space by s bit-level importance coding. 

\textbf{Hashing-based neural CF} 
Recent hashing-based CF methods utilize neural networks to extract user/item features (representation). Autoencoders (AEs) \cite{tagaware}, Variational Autoencoders (VAEs)\cite{neuhash,pqvae}, and Graph neural networks   were used for feature extraction\cite{stenew,10.1145/3442381.3449884}.  Zhang \etal \cite{nbr} introduced auto-encoders to learn continuous features from side information (e.g. user demographics and item information). These values were converted into binary codes with bit-by-bit optimization similar to DCF~\cite{discrete} and were used as an input for collaborative filtering algorithms. In terms of interaction modeling, cosine similarity loss\cite{tagaware} and rank loss\cite{fashiontanh} were studied in addition to traditional rating reconstruction loss. Several binarization techniques were also introduced. HashNet \cite{hashnet} used a scaled $tanh$function which forces real-valued features  close to $+1$ and $-1$ during training. Others used straight-through-estimator (STE) to solve the discrete optimization problem directly\cite{ungsh, nash,stenew}. 

\textbf{Hashing-based similarity-preserving cross-modal retrieval}   
Creating compact codes for cross-modal retrieval has been intensively explored to satisfy the needs from the massive growth of multi-modal data\cite{cmssh,cmfh,seph}. 
The state-of-the-art compact cross-modal retrieval methods utilize neural networks to extract real-valued features and convert them to binary codes use binarizations such as $sign$ and scaled $tanh$ functions  \cite{dcmh,cmhh,djsrh,10.1145/3404835.3462888}. The main focus of these methods is defining better objective functions that can express similarities between entities from different modalities. Joint-
Modal Distribution-Based Similarity Hashing (JSDH)~\cite{jdsh} constructed a joint-modal similarity matrix to  preserve the cross-modal semantic correlations among instances. Unsupervised Generative Adversarial Cross-modal Hashing (UGACH)~\cite{ugach} proposed to use a generative adversarial network  to learn better underlying features from multiple modalities. High dimensional sparse cross-modal hashing (HSCH)~\cite{10.1145/3442381.3449798} studied High-dimensional Sparse Hashing for cross-modal retrieval that maps inputs into a higher dimensional space and generates sparse hash codes instead of the traditional dense hashing.   Multi-Index Semantic Hashing (MISH) ~\cite{10.1145/3442381.3450014} introduced multi-index semantic hashing to improve searching efficiency over binary hashing codes.


\section{Notations}
Assume we have $m$ users, $n$ items, and an implicit feedback matrix $\mathbf{S} \in \R^{m \times n}$ which contains binary values 0 or 1. $\mathbf{S}_{ab} = 1$ means User $a$ likes Item $b$, $\mathbf{S}_{ab} = 0$ means User $a$ dislikes Item $b$ or has an unknown preference about Item $b$. Note that an explicit feedback matrix $\mathbf{R}\in \R^{m \times n}$ can be converted to $\mathbf{S}$ by setting a threshold (i.e., if $\mathbf{R}_{ab}>3$, $\mathbf{S}_{ab} =1$, 0 otherwise). 
Given $\mathbf{S}$, we want to learn real-valued feature matrices $\mathbf{F}^u \in \R^{m \times r}$, $\mathbf{F}^i \in \R^{n \times r}$,  and binary feature matrices $\mathbf{H}^u \in \{-1,+1\}^{m \times r}$, $\mathbf{H}^i \in \{-1,+1\}^{n \times r}$  for users and items. $\mathbf{F}^u_a$ is User $a$’s continuous features (i.e., real-valued representations) and  $\mathbf{F}^u_b$ is Item $b$’s continuous features. All users and items are represented by $r$-dimensional vectors.
\section{Compact Cross-Similarity Recommendation}
\label{sec:crr}
In this section, we detail the modules and the final objective function of CCSR. As we described in Section\ref{sec:intro}, hashing-based RSs consist of three modules: feature extraction, interaction modeling, and binarization. In this Section, we demonstrate our design choice for each module and the difference from previous work. 
\subsection{Feature Extraction}

We use autoencoders to learn continuous features and convert them into binary codes, following~\cite{neuhash, tagaware}. We do not use multilayer perceptron as previous literature showed that  autoencoders converge faster than multilayer perceptrons~\cite{tagaware}. We construct two autoencoders, one for users and the other for items. The inputs are user and item low-level features. It can be purchase history, ratings, and side information such as user's demographics or item titles. Here we use a rating  matrix as an input:  $\mathbf{X} = \mathbf{R}$ for user, and  $\mathbf{Y} = \mathbf{R}^\top$ for items. The autoencoders convert the original input $\mathbf{X}_a$ and $\mathbf{Y}_b$ into low-dimensional representations $\mathbf{F}^u_a$ and $\mathbf{F}^i_b$, and produce reconstructed input $\hat{\mathbf{X}}_a$ and $\hat{\mathbf{Y}}_b$ from these low-dimensional representations. The loss function for the autoencoders is: 
\begin{equation}
  \mathcal{L}_{ae} = \sum_{a,b}( || \mathbf{X}_a -\mathbf{\hat{X}}_a ||_F^2  +|| \mathbf{Y}_b -\mathbf{\hat{Y}}_b ||_F^2  ),
\end{equation} 
\noindent where $\mathbf{\hat{X}}_a$, $\mathbf{\hat{Y}}_b$ are outputs of the autoencoders. Low-dimensional continuous feature matrices $\mathbf{F}_a^u$, $\mathbf{F}_b^i$ are the outputs of each encoder's last layer.

\subsection{User-item Interaction}
\label{sec:loss}
The user-item interaction loss function of CCSR consists of two parts: 1) similarity loss and 2) balance loss. Let's assume that we have two data points $a,b$ that come from different modalities (i.e., users and items). They have continuous features $\mathbf{F}_a^u$, $\mathbf{F}_b^i$, as shown in Figure~\ref{fig:model}. We define similarity label $\mathbf{S}_{ab}$. $\mathbf{S}_{ab}=1$ implies $a, b$ are similar, whereas $\mathbf{S}_{ab} = 0 $ implies they are dissimilar.

\textbf{Similarity loss}
To measure the similarity between entities from different modalities, we adopt the Maximum a Posteriori (MAP) estimation. 
The logarithm MAP is:
\begin{equation}
\log p(\mathbf{F}_a^u,\mathbf{F}_b^i|\mathbf{S}_{ab}) \propto \log p(\mathbf{S}_{ab}|\mathbf{F}_a^u,\mathbf{F}_b^i)p(\mathbf{F}_a^u)p(\mathbf{F}_b^i) \label{eqn:map}
\end{equation}
In Equation~\ref{eqn:map}, 
The conditional likelihood for similarity label $S_{ab}$ is:
\begin{equation}
\begin{aligned}
  p(\mathbf{S}_{ab}|\mathbf{F}_a^u,\mathbf{F}_b^i) &=
    \begin{cases}
      \sigma(\langle \mathbf{F}_a^u,\mathbf{F}_b^i\rangle) & \mathbf{S}_{ab} = 1\\
      1-\sigma(\langle \mathbf{F}_a^u,\mathbf{F}_b^i\rangle) & \mathbf{S}_{ab} = 0
    \end{cases} \\    
&= \sigma(\langle \mathbf{F}_a^u,\mathbf{F}_b^i\rangle)^{\mathbf{S}_{ab}}(1-\sigma(\langle \mathbf{F}_a^u,\mathbf{F}_b^i\rangle))^{1-\mathbf{S}_{ab}}
\end{aligned}
\label{eqn:condition1}
\end{equation}
where $\sigma(x) = 1/(1+e^{-x})$ is the sigmoid function, $\langle \rangle$ is the inner product operator. Assuming the priors for $\mathbf{F}_a^u$ and $\mathbf{F}_b^i$ are known and follow Gaussian distribution,
\begin{equation}
    p(\mathbf{F}_a^u) = \frac{1}{\sqrt{2\pi} \sigma}exp(-|| \mathbf{F}_a^u- \mathbf{\hat{F}}_a^u||^2/2 \sigma^2)
 \label{eqn:condition2}   
\end{equation}
\begin{equation}
    p(\mathbf{F}_b^i) = \frac{1}{\sqrt{2\pi} \sigma}exp(-|| \mathbf{F}_b^i- \mathbf{\hat{F}}_b^i||^2/2 \sigma^2)
    \label{eqn:condition3}
\end{equation}
By maximizing Equation~\ref{eqn:map} with Equation~\ref{eqn:condition1},~\ref{eqn:condition2},~\ref{eqn:condition3}, we obtain the cross-entropy loss for MAP-based similarity:
\begin{equation}
\mathcal{L}_{sim} = \sum_{{a,b}} (\log(1+e^{\langle \mathbf{F}_a^u,\mathbf{F}_b^i\rangle })-  \mathbf{S}_{ab}\langle \mathbf{F}_a^u,\mathbf{F}_b^i\rangle  ).
 \label{eqn:simloss}
\end{equation} 
Note that previous recommender system models estimate the ratings using dot-product similarity between users and items. Therefore, their loss function is as follows:
$\mathcal{L}^{'}_{sim} = \sum_{{a,b}} (\mathbf{R}_{ab}-{\langle \mathbf{F}_a^u,\mathbf{F}_b^i\rangle })^2$.

\textbf{Balance loss} 
To ensure we use the bit information maximally, we utilize a balance loss to balances the number of $+1$ and $-1$ in the binary code, which is a widely used technique in binary code learning~\cite{cmfh,NIPS2008_d58072be}.  Since it is difficult to directly optimize on binary codes, we apply the balance loss to the continuous features: 
$
   \mathcal{L}_b = \sum_{a,b} ( ||\mathbf{F}_a\mathbf{1}||_F^2+ ||\mathbf{F}_b\mathbf{1}||_F^2 )
$
where $\mathbf{1}$ is a vector of 1s.  

As a result, the loss function of CCSR is defined as:
\begin{equation}
    \mathcal{L} = \mathcal{L}_{sim}+\lambda_b \mathcal{L}_b + \lambda_{ae} \mathcal{L}_{ae}
    \label{eqn:loss1}
\end{equation}
\noindent where $ \lambda_{ae}$ and $\lambda_b$ are the hyper-parameters to balance between the losses. 
 \subsection{Binarization}
 \label{sec:learnbin}
 In the previous section, we optimize the model over continuous features. The binarization can be done by applying a $sign$  function to the continuous feature:
\begin{equation}
\begin{aligned}
  h = sgn(f)  &=
    \begin{cases}
      1 & f \geq 0\\
      -1 & f < 0 ,
    \end{cases} \\    
\end{aligned}
\label{eqn:binary}
\end{equation}
\noindent which is a two-stage binarization. In~\cite{hashnet,djsrh}, authors replaced Eq.~\ref{eqn:binary} with a scaled $tanh$ function: 
$
 h = tanh(\alpha f)   
$.
Starting with $\alpha =1$, the method increases  $\alpha$ exponentially per training epoch so that eventually $tanh(\alpha f) \approx sgn(f)$. This replacement aim to learnining binary codes by directly optimizing the neural networks (i.e., end-to-end binarization). However, this approximation still outputs \emph{continuous} values. We apply a $sign$ function to output of $tanh(\alpha f)$  to get true binary codes. We call this method sign-scaled $tanh$.

\section{Experiments}
In this section, we perform experiments to answer the following questions:
\begin{itemize}
    \item[\textbf{Q1}] What is the performance difference when using Matrix factorization-based (MF-based) and Cross-modal retrieval-based (CR-based)  similarity measurements?
    \item[\textbf{Q2}] What is the performance difference between different binarization methods: sign (S), scaled $tanh$ (ST), and sign-scaled $tanh$ (SST)?
    \item[\textbf{Q3}] When similarity measurement is more effective than MF-based measurement?
\end{itemize}
\subsection{Dataset}
We use three datasets for the experiments: Movielens1M~\cite{ml100k},  Amazon (books) dataset~\cite{amazon} and a real-word e-commerce ( Ichiba\footnote{https://rit.rakuten.com/data\_release/}) purchase dataset. Movelens1M and Amazon datasets contain explicit feedback; they contain movie and book ratings from users. Ratings vary from 1 to 5 and unrated items have 0 ratings. Ichiba dataset contains implicit feedback; it contains purchase histories in one category.

\textbf{Pre-processing} We utilize  Movielens1M without any user/item filtering. Following previous work~\cite{neuhash}, for Amazon dataset,  we remove users and items with insufficient ratings, and as a result, all users and items in the dataset have at least 20 ratings. To generate the similarity matrix, ratings greater than 3 are converted to 1, and the rest to 0.  For Ichiba dataset, we filter out users and items with insufficient interactions so that all users and items have at least 50 interactions. We assume that if user $a$ purchases item $b$, then $a$,$b$ are similar.

\textbf{Training/test split}  For Movielens1M, we followed previous work~\cite{bccf}. \emph{For each user},   randomly select 80\% ratings for training and 20\% for the test. For Amazon and Ichiba, since they are extremely sparse, \emph{for all ratings}, we randomly select 80\%  for training and 20\% for the test. A detailed datasets summary is listed in Table~\ref{tab:dataset}.


\begin{table}[!t]
  \caption{Details of the datasets}
  \label{tab:dataset}
  \centering
  \scalebox{1}{
  \begin{tabular}{ccccc}
    \toprule
    Dataset& \#User & \#Item & \#Ratings & Density\\
    \midrule
Movielens1M& 6,040 & 3,952 & 1,000,209 &4.19\% \\
Amazon & 35,736 & 38,121 & 1,960,674 & 0.14\% \\
Ichiba & 36,314 & 8,514 & 1,267,296 &0.41\%  \\
  \bottomrule
\end{tabular}
}
\end{table}

\subsection{Models}
\label{sec:models}
We compared the recommendation performance of the following models:
\begin{enumerate}
    \item \textbf{Random} We randomly select  items from all items and recommend them to each test user.
    \item \textbf{Top} We select the most frequently rated items from training data and recommend them to all test users.
    \item \textbf{CF} 
    This is a Matrix Factorization method based on Stochastic Gradient Descend. 
    The feature matrices $\mathbf{F}^u, \mathbf{F}^i$ for users and items are randomly initialized and updated alternatively through gradient descent over the following loss function: 
    \begin{equation}
        \sum_{a,b} (\mathbf{R}_{ab}-<\mathbf{F}^u_a, \mathbf{F}^i_b>)^2  + \lambda ||\mathbf{F}^u||_F^2 + \lambda ||\mathbf{F}^i||_F^2.
    \end{equation}
    
    Continuous features are converted to binary codes using $sign$ function. 
    \item \textbf{CFcodeReg}~\cite{bccf} This method first solves a relaxed optimization problem where features can be continuous and between -1 and 1. 
    The loss function is: 
    \begin{equation}
        \sum_{a,b} (\mathbf{R}_{ab}-\frac{1}{2}-\frac{1}{2r}<\mathbf{F}^u_a, \mathbf{F}^i_b>)^2  + \lambda ||\mathbf{F}^u||_F^2 + \lambda ||\mathbf{F}^i||_F^2,   
    \end{equation}
    where $r$ is the dimension of the continuous features. All features are rounded to the closest binary code with median threshold. 
    \item \textbf{AECF} 
    We use autoencoders introduced in Section~\ref{sec:crr} with a rating reconstruction loss. The hidden-layer sizes are 512, 256 and 128 for the three-layer encoder and 128, 256, and 512 for the decoder. The number of layers and the size of layers are set after cross-validation. The loss function is:
    \begin{equation}
    \sum_{a,b}( (\mathbf{R}_{ab}-<\mathbf{F}^u_a, \mathbf{F}^i_b>)^2 + \lambda_{ae} \mathcal{L}_{ae}    
    \end{equation}
    Depending on the binarization methods, we have \textbf{AECF-S} with a $sign$ function, \textbf{AECF-ST} with a scaled $tanh$ function, and \textbf{AECF-SST} with a sign-scaled $tanh$ function.
    \item \textbf{DJSRH~\cite{djsrh}} Deep joint-semantics reconstructing hashing (DJSRH) is the state-of-the-art compact cross-modal retrieval model. It uses semantic loss in both single-modality and cross-modality for better performance. The features come from a multilayer perceptron (MLP) with two hidden layers whose sizes are 512 and 256. It utilizes an unsupervised similarity loss: 
    \begin{equation}
        ||\mu \mathbf{M} -\mathbf{F}^u\mathbf{F}^{i\top}||_\mathbf{F}^2,
    \end{equation}
    where M is a joint semantics affinity matrix that includes both single and cross modality similarities ($\mathbf{M} = (1-\eta)\mathbf{N}+\eta \mathbf{N}\mathbf{N}^\top, \mathbf{N} = \beta \mathbf{X}\mathbf{X}^\top+(1-\beta) \mathbf{Y}\mathbf{Y}^\top$). $\eta$ and $\beta$ are parameters to balance between different terms. We have \textbf{DJSRH-ST} and \textbf{DJSRH-SST}.

    \item \textbf{CCSR}  Our proposed method. 
    The loss function is Equation~\ref{eqn:loss1}. We choose a one-layer encoder/decoder using cross-validation.
    The size of the hidden-layer is 128 for the encoder/decoder. We also investigate different binarization methods: \textbf{CCSR-S}, \textbf{CCSR-ST}, and \textbf{CCSR-SST}.
\end{enumerate}

The first two models are rule-based. The following four methods are MF-based and the last two are CR-based. Besides binarized AECF and CCSR, we also performed the  AECF and CCSR with continuous-values, noted as \textbf{AECF-C} and \textbf{CCSR-C}, for reference. Since the paper focuses on binary coding, AECF-C and CCSR-C are not discussed in Q1 and Q2.

We trained and tuned DJSRH  using the code provided by the authors\footnote{https://github.com/zzs1994/DJSRH}. We carefully implemented and tuned the rest of the models. In CF and CFcodeReg, we set $\lambda = 0.4$.
In AECF and CCSR models, we cross-validated the hyper parameters and finally set  $\lambda_{ae}$ = 0.1, $\lambda_{b}$ = 0.0001 for Movielens1M and $\lambda_{ae}$ = 10, $\lambda_{b}$ = 0.0001 for Amazon and Ichiba. We added one dropout layer before last encoder layer in both AECF and CCSR. Dropout rate was set at 0.6 for code length 5, 10 and 20 experiments, and 0.8 for code length 40 through cross-validation. 

\begin{table*}[h]
  \caption{NDCG@k  of different models on Movielens1M ( Final binary code lengths are 5,10,20,40.)}
  \label{tab:ml1mndcg}
  \centering
  \scalebox{1}{
  \begin{tabular}{c|llll|llll|llll}
    \toprule
    \multirow{2}{*}{\textbf{Models}}& \multicolumn{4}{c|}{@2} & \multicolumn{4}{c|}{@6} &  \multicolumn{4}{c}{@10} \\
    \cline{2-13}
    & \multicolumn{1}{c}{5} & \multicolumn{1}{c}{10}& \multicolumn{1}{c}{20} &\multicolumn{1}{c|}{40} & \multicolumn{1}{c}{5} & \multicolumn{1}{c}{10}& \multicolumn{1}{c}{20} &\multicolumn{1}{c|}{40} & \multicolumn{1}{c}{5} & \multicolumn{1}{c}{10}& \multicolumn{1}{c}{20} &\multicolumn{1}{c}{40} \\
    \midrule
    
Random &\multicolumn{4}{c|}{0.0094} & \multicolumn{4}{c|}{0.0091 }& \multicolumn{4}{c}{0.0083} \\
Top & \multicolumn{4}{c|}{0.0}& \multicolumn{4}{c|}{0.3101} &\multicolumn{4}{c}{ 0.3196}\\
\hline
AECF-C & 0.6496 &0.7093  &0.6750  &0.6694 & 0.7072 & 0.7358 &0.7129 &0.7100 & 0.7420  & 0.7603 &0.7424 & 0.7393\\
CCSR-C & 0.6861 & 0.7010 &0.6901& 0.7246 &0.7296  &0.7344 &0.7255  & 0.7605&0.7538  & 0.7605 &0.7516 &0.7885  \\
\hline
  AECF-ST &  0.6201  & 0.6862  & \textbf{0.6769}& \textbf{0.6918}&0.6762  & 0.7220 & \textbf{0.7180} & \textbf{0.7246} &0.7095  & 0.7495  & \textbf{0.7458}&\textbf{0.7518} \\
    DJSRH-ST& 0.5661 &0.6020 &0.5974&0.5879 &0.6274 &0.6533 &0.6481&0.6409 &0.6673& 0.6886 & 0.6845 & 0.6787  \\
    CCSR-ST & \textbf{0.6820} &\textbf{0.6913 } &0.6441 &0.6610 & \textbf{0.7226}  &\textbf{0.7295} &0.6841  &0.7016 & \textbf{0.7470} & \textbf{0.7542 }& 0.7115 &0.7291 \\
      \hline
    AECF-SST & 0.5637 & \textbf{0.6798}  & \textbf{0.6883}  &0.5404 &0.6233 &\textbf{0.7201}& \textbf{0.7265} &0.6032&0.6663  &\textbf{0.7478} & \textbf{0.7526} &0.6445\\
   DJSRH-SST& 0.5676&0.5825 &0.5895&0.5859&0.6319 &0.6388 &0.6458&0.6446&0.6703&0.6776 &0.6835 &0.6814\\
    CCSR-SST & \textbf{0.6492} &0.6668  &0.6587 &\textbf{0.6609}& \textbf{0.6930} &0.7077 & 0.7108 & \textbf{0.7138} & \textbf{0.7238} &0.7384 &0.7438& \textbf{0.7486 } \\
    \hline    
CF & 0.5492&0.5599  &0.5672&0.5833&0.6172  &0.6198 &0.6297&0.6450 & 0.6593  & 0.6613 & 0.6698 & 0.6840 \\
CFCodeReg &0.5692  &0.5690  &0.5738 &0.5728&0.6314 &0.6303 &0.6323 &0.6317&0.6712& 0.6701 & 0.6724 & 0.6711\\
    AECF-S & 0.4983 &0.5555&0.5310  &0.4874 &0.5817 &0.6154 &0.5989&0.5689 &0.6311 & 0.6534 & 0.6423  & 0.6175  \\
    CCSR-S & \textbf{0.6332} &\textbf{0.6523}  &\textbf{0.6912 } &\textbf{0.7402} & \textbf{0.6997 } & \textbf{0.7128} &\textbf{0.7277}&\textbf{0.7677}& \textbf{0.7371} &\textbf{0.7475}&\textbf{0.7529 }& \textbf{0.7897 } \\

  \bottomrule
\end{tabular}
}
\end{table*}

\begin{table*}[h]
  \caption{NDCG@k  of different models on Amazon} 
  \label{tab:amazonndcg}
  \centering
  \scalebox{1}{
  \begin{tabular}{c|llll|llll|llll}
    \toprule
    \multirow{2}{*}{\textbf{Models}}& \multicolumn{4}{c|}{@2} & \multicolumn{4}{c|}{@6} &  \multicolumn{4}{c}{@10} \\
    \cline{2-13}
    & \multicolumn{1}{c}{5} & \multicolumn{1}{c}{10}& \multicolumn{1}{c}{20} &\multicolumn{1}{c|}{40} & \multicolumn{1}{c}{5} & \multicolumn{1}{c}{10}& \multicolumn{1}{c}{20} &\multicolumn{1}{c|}{40} & \multicolumn{1}{c}{5} & \multicolumn{1}{c}{10}& \multicolumn{1}{c}{20} &\multicolumn{1}{c}{40} \\
    \midrule

Random &\multicolumn{4}{c|}{0.0003} & \multicolumn{4}{c|}{ 0.0003}& \multicolumn{4}{c}{0.0001} \\
Top & \multicolumn{4}{c|}{0.0132}& \multicolumn{4}{c|}{0.0327} &\multicolumn{4}{c}{0.0485}\\
 \hline
   AECF-C &0.7919  &0.7915&0.7944
&0.7930&0.8563&0.8558 &0.8578&0.8566&0.8822 & 0.8816 & 0.8834 & 0.8824 \\
   CCSR-C & 0.7875 &
   0.7883&0.7880 &0.7895 &
 0.8526 & 0.8522&0.8526 &0.8539&0.8791 & 0.8791 & 0.8792 & 0.8801 \\
     \hline
   AECF-ST &  0.7800  &0.7716 &0.7810   &0.7835     & 0.8471  & 0.8419  &0.8478&0.8489  &  0.8748   &0.8705&0.8752 &0.8765  \\
   DJSRH-ST&  0.7460 &0.7653 &0.7707  &0.7792  & 0.8263 &0.8381&0.8412&0.8469 & 0.8579  &0.8673 &0.8701 &0.8744  \\
       CCSR-ST &  \textbf{0.7833} &\textbf{0.7869} &\textbf{0.7837} &\textbf{0.7870} & \textbf{0.8490} &\textbf{0.8521} &\textbf{0.8503} &\textbf{0.8530}& \textbf{0.8762}  &\textbf{0.8786} &\textbf{0.8770} &\textbf{0.8792 } \\

      \hline

   AECF-SST &  \textbf{0.7657} &0.7500  &\textbf{0.7817}  &\textbf{0.7733} & \textbf{0.8377 } &0.8288& \textbf{0.8485} &\textbf{0.8432}& \textbf{0.8670} &0.8599&\textbf{0.8759} &\textbf{0.8714}\\
   DJSRH-SST&  0.7610  &0.7617 &0.7627  &
   0.7698 & 0.8356 &\textbf{0.8372} &0.8379 &0.8421& 0.8654 &\textbf{0.8668} & 0.8676 &0.8709\\

    CCSR-SST &  0.7632 & \textbf{0.7633}  &0.7518 &0.7530  & 0.8355  &0.8362&0.8285&0.8306& 0.8657 &0.8658 &0.8598 &0.8614 \\

\hline
CF &  0.7661  &0.7673 &0.7680 &0.7691 & 0.8399  &0.8409 &0.8422 & 0.8428 & 0.8692 & 0.8700  &0.8711 &  0.8716\\
CFCodeReg & 0.7657  &
0.7650  &0.7653 &0.7663 & 0.8398  &0.8400 &0.8405&0.8413 &0.8692 & 0.8693 &  0.8698& 0.8704  \\

     AECF-S &  0.7703 & \textbf{0.7755} &\textbf{0.7818}  &\textbf{0.7793}  & 0.8411 &\textbf{0.8447 } &\textbf{0.8481}&\textbf{0.8477} & 0.8697 &\textbf{0.8726}  & \textbf{0.8756} & \textbf{0.8750}  \\
         CCSR-S & \textbf{ 0.7707}   &0.7685 &0.7648 &0.7752 & \textbf{0.8420} &0.8408 &0.8381 &0.8446 & \textbf{0.8705}  &0.8697 &0.8676 &0.8725 \\

  
  \bottomrule
\end{tabular}
}
\end{table*}

\begin{table*}[!t]
  \caption{NDCG@k  of different models on Ichiba}
  \label{tab:ichibandcg}
  \centering
  \scalebox{1}{
  \begin{tabular}{c|llll|llll|llll}
    \toprule
    \multirow{2}{*}{\textbf{Models}}& \multicolumn{4}{c|}{@2} & \multicolumn{4}{c|}{@6} &  \multicolumn{4}{c}{@10} \\
    \cline{2-13}
    & \multicolumn{1}{c}{5} & \multicolumn{1}{c}{10}& \multicolumn{1}{c}{20} &\multicolumn{1}{c|}{40} & \multicolumn{1}{c}{5} & \multicolumn{1}{c}{10}& \multicolumn{1}{c}{20} &\multicolumn{1}{c|}{40} & \multicolumn{1}{c}{5} & \multicolumn{1}{c}{10}& \multicolumn{1}{c}{20} &\multicolumn{1}{c}{40} \\
    \midrule

Random &\multicolumn{4}{c|}{0.0010 } & \multicolumn{4}{c|}{0.0013 }& \multicolumn{4}{c}{0.0009} \\
Top & \multicolumn{4}{c|}{0.0558}& \multicolumn{4}{c|}{0.1113} &\multicolumn{4}{c}{0.1550}\\
\hline
AECF-C &0.9180& 0.9233& 0.9199&
0.9168 &0.9508 &0.9537 &0.9506 &0.9479&0.9606 &0.9628  &0.9604 &0.9581 \\
CCSR-C & 0.9216 &  0.9236  & 0.9196&0.9293&0.9482& 0.9567 &0.9561&0.9531&0.9593&0.9607 &0.9677&0.9625 \\
\hline
      AECF-ST &  
             0.9062 
               &0.9141  &0.8955 &0.9108  & 0.9386 &0.9473  &0.9387 &0.9474 &0.9512  &0.9574 &0.9513 &0.9578  \\ 
           DJSRH-ST& 0.8730 &\textbf{0.9202} &
           \textbf{0.9406}& \textbf{0.9656} &0.9196  &\textbf{0.9491} &\textbf{0.9593}&\textbf{0.9730} &0.9385   &\textbf{0.9597} &\textbf{0.9674}&\textbf{0.9784} \\
               CCSR-ST & \textbf{0.9095} &
    0.9166 
& 0.9124  &  0.9227 & \textbf{0.9437}  & 0.9487 & 0.9463 & 0.9526& \textbf{0.9551}  & 0.9588& 0.9572& 0.9617 \\

 \hline
   
         AECF-SST &  0.9020 &0.8951 &0.9016  &0.9072  & 0.9377 &0.9389&0.9365&0.9398&  0.9510 &0.9511 &0.9500&0.9523 \\
         DJSRH-SST& 0.8954 &\textbf{0.9119 } &\textbf{0.9188} &\textbf{0.9434}  &0.9365 &\textbf{0.9464} &\textbf{0.9497}&\textbf{0.9629}& 0.9500  &\textbf{0.9577}&\textbf{0.9602}&\textbf{0.9705} \\
    CCSR-SST &  \textbf{0.9145 } &0.9066  &0.8965   &0.8695 & \textbf{0.9449}&0.9384&0.9317 &0.9187& \textbf{0.9562} &0.9512 &0.9464&0.9369  \\
      \hline  
CF &  0.8915 &0.8921 &0.8927 &0.8956 &0.9329 &0.9339& 0.9338  &0.9352 &0.9475 & 0.9482  &0.9483 &0.9494\\
CFCodeReg & 0.8913  &0.8911 &0.8875  &0.8856 & 0.9327  &0.9322 &0.9307 &0.9300 & 0.9475  &0.9470 &0.9457 &0.9452 \\
    AECF-S &  0.9150 &0.9102 & \textbf{0.9123}  &
    0.9123  
    & 0.9454 & 0.9416  & \textbf{0.9430}   &0.9426 & 0.9566  & 0.9532 & \textbf{0.9547}  &0.9542 \\
            
             CCSR-S &  \textbf{0.9169}  &\textbf{0.9104}  & 0.9026  & \textbf{0.9158}  & \textbf{0.9467}  &\textbf{0.9432} &0.9384  &\textbf{0.9478} &  \textbf{0.9576}  &\textbf{0.9548} & 0.9513 & \textbf{0.9583}\\

  \bottomrule
\end{tabular}
}
\end{table*}

\begin{table*}
\centering
\caption{Recall@6 on three datasets}
\label{tab:recall}
\begin{tabular}{c|llll|llll|llll} 
\toprule
\multirow{2}{*}{\textbf{Models}} & \multicolumn{4}{c|}{\textbf{MovieLen1M }} & \multicolumn{4}{c|}{\textbf{Amazon }} & \multicolumn{4}{c}{\textbf{Ichiba }} \\ 
\cline{2-13}
 & \multicolumn{1}{c}{5} & \multicolumn{1}{c}{10} & \multicolumn{1}{c}{20} & \multicolumn{1}{c|}{40} & \multicolumn{1}{c}{5} & \multicolumn{1}{c}{10} & \multicolumn{1}{c}{20} & \multicolumn{1}{c|}{40} & \multicolumn{1}{c}{5} & \multicolumn{1}{c}{10} & \multicolumn{1}{c}{20} & \multicolumn{1}{c}{40} \\ 
\midrule
Random & \multicolumn{4}{c|}{0.0260} & \multicolumn{4}{c|}{0.0001} & \multicolumn{4}{c}{0.0007} \\
Top & \multicolumn{4}{c|}{0.0282} & \multicolumn{4}{c|}{0.0045} & \multicolumn{4}{c}{0.0167} \\ 
\hline
AECF-ST & 0.4384 & 0.4497 & 0.4511 & 0.4561 & 0.7363 & 0.7364 & 0.7367 & 0.7355 & 0.8012 & 0.8103 & 0.8058 & 0.8120 \\
DJSRH-ST & 0.4184 & 0.4260 & 0.4233 & 0.4183 & 0.7350 & 0.7360 & 0.7364 & 0.7375 & 0.7890 & 0.8107 & \textbf{0.8165 } & \textbf{0.8249 } \\
CCSR-ST & \textbf{0.4671} & \textbf{0.4660} & \textbf{0.4560} & \textbf{0.4599} & \textbf{0.7367} & \textbf{0.7369 } & \textbf{0.7367 } & \textbf{0.7375 } & \textbf{0.8012} & \textbf{0.8107 } & 0.8093 & 0.8146 \\ 
\hline
AECF-SST  & 0.4181 & 0.4522 & 0.4552 & 0.4121 & 0.7348 & 0.7343 & \textbf{0.7365 } & 0.7358 & 0.7951 & 0.7972 & 0.7887 & 0.7957 \\
DJSRH-SST & 0.4187 & 0.4244 & 0.4275 & 0.4275 & \textbf{0.7352 } & \textbf{0.7357 } & 0.7361 & \textbf{0.7370 } & 0.7948 & \textbf{0.8057} & \textbf{0.8089} & \textbf{0.8190} \\
CCSR-SST & \textbf{0.4524 } & \textbf{0.4593 } & \textbf{0.4588 } & \textbf{0.4579 } & 0.7347 & 0.7351 & 0.7344 & 0.7349 & \textbf{0.7951} & 0.8000 & 0.8027 & 0.8086 \\ 
\hline
CF & 0.4159 & 0.4173 & 0.4209 & 0.4268 & 0.7359 & 0.7361 & 0.7360 & 0.7360 & 0.7949 & 0.7999 & 0.7983 & 0.7996 \\
CFCodeReg & 0.4200 & 0.4206 & 0.4215 & 0.4191 & 0.7357 & 0.7357 & 0.7358 & 0.7358 & 0.7943 & 0.7953 & 0.7958 & 0.7963 \\
AECF-S & 0.4015 & 0.4226 & 0.4201 & 0.4040 & 0.7349 & 0.7357 & \textbf{0.7363 } & 0.7360 & 0.7859 & 0.7889 & 0.7866 & 0.7889 \\
CCSR-S & \textbf{0.4544 } & \textbf{0.4515 } & \textbf{0.4522 } & \textbf{0.4697 } & \textbf{0.7357} & \textbf{0.7360} & 0.7359 & \textbf{0.7361 } & \textbf{0.7998} & \textbf{0.8012} & \textbf{0.7997} & \textbf{0.8080 } \\
\bottomrule
\end{tabular}
\end{table*}

\subsection{Evaluation Metrics}
We use Recall@k and Normalized Discounted Cumulative Gain (NDCG)@k as the evaluation matrices. We recommend the top-k items to a query user. The top-k items are ranked based on the Hamming distance between binary codes of a query user and the items in the database. The true labels come from similarity/rating matrix. 

\subsection{CR-based Recommender v.s. MF-based Recommender (Q1)}
In Table~\ref{tab:ml1mndcg},~\ref{tab:amazonndcg}, and ~\ref{tab:ichibandcg} we show NDCG results on Movielens1M, Amazon, and Ichiba. CCSR-S models achieve the best performance across different binary code length in all models with $sign$ binarization on Movielens1M. It has an average 15.21\% NDCG relative improvement compared to the best MF-based models. On Amazon dataset, CCSR-S performed as good as the best MF-based model and CCSR-ST outperforms all MF-based models. 
On the Ichiba purchase dataset, the CR-based model DJSRH performs best, followed by the CCSR model. Overall, DJSRH-SST  achieves  average 1.26\% NDCG relative improvement compared to the best MF-based model with the same binarization method. We see similar results in Recall evaluations in Table~\ref{tab:recall}.  

We conclude that CR-based similarity loss is better than MF-based loss in  the top-k recommendation task. This is because the former loss emphasizes more on similarity learning which is more important for the recommendation task. The latter focuses on rating reconstruction which is an indirect approach to the task and may cause over-fitting. While performance still increases when we increase feature size from 20 to 40 in CCSR models, AECF models suffer from performance drop. We present a further analysis in Section \ref{sec:two_losses}.

\textit{Among MF-based models}, AECF achieves the best performance because it uses neural feature extractors.  CFcodeReg and CF have similar performances which indicate median threshold and sign threshold have little influence on performance.

\textit{Among CR-based models}, CCSR performs better than DJSRH on Movielens1M. This is reasonable since unsupervised similarity measurements in DJSRH are less powerful than supervised measurements like CCSR when data density is relatively high. However, on Ichiba dataset, DJSRH achieves better performances. Besides the effects of low data density, the low diversity in input features (0/1 in purchase data and 1 to 5 in rating data) in Ichiba also makes autoencoder-based CCSR less competitive. On the other hand, DJSRH can learn better similarity relations in the large, sparse dataset due to the intra and inter similarity loss measurements in the algorithm.
Also, it is notable that the Top item recommendation shows decent performance in NDCG evaluation compared to the Random model, as we see in Table~\ref{tab:ml1mndcg},~\ref{tab:amazonndcg},~\ref{tab:ichibandcg}. We see similar observations in~\cite{rich}. 

\subsection{Different Binarization  Methods (Q2)}
\begin{figure}
    \centering
    \includegraphics[width=8cm]{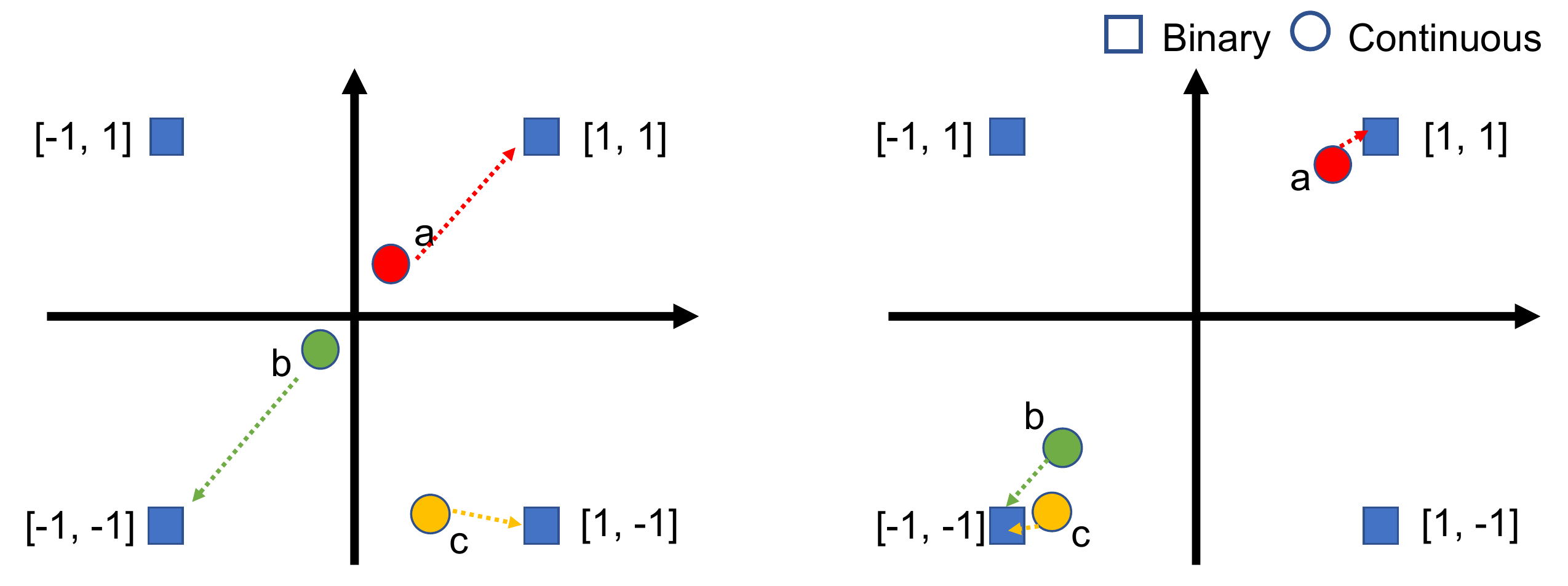}
    \caption{A toy example to explain reasons for performance differences using continuous features and various binary features (ST and SST). General limitations of binary features (on the left): even though $a$ is more similar to $b$ than to $c$ in continuous space, after being converted to binary codes, $a$ is more similar to $c$ than to $b$ when comparing Hamming distance. Limitations of SST (on the right): In ST models, continuous features are close to +1 and -1. $a$ is more similar to $b$ than to $c$, but $b$ and $c$ are equally similar to $a$ after converting use SST.}
    \label{fig:limit_b}
\end{figure}

As we discussed in Section~\ref{sec:learnbin}, there are three ways to learn the binary codes: S, ST, and SST.
Recent papers prefer the scaled $tanh$ (ST) function to $sign$ (S) function because ST can be computed in the back-propagation step and learn binary codes in the end-to-end fashion. Technically, however, the output of ST is not binary code, so it is unfair to compare the performance of ST with S or SST whose output is binary. For a fair comparison, we applied a $sign$ function to the output of ST and made them binary (SST).
\textit{We find the performance drops drastically when we switch from ST to SST.} From Table~\ref{tab:ml1mndcg},\ref{tab:recall}, for each method, NDCG/Recall value decreases more than 1\% when the output of ST was forced to be binary (SST). We explain the possible reasons in Figure~\ref{fig:limit_b}. In summary, the performance change is because of the limited representation power of binary codes.


The drop by SST in AECF and CCSR is higher than the drop in DJSRH. This might be due to the different focuses in the similarity loss; DJSRH compares the similarities in a batch (rank loss  between different users and items), so it learns robust codes. AECF and CCSR only consider pair-wise loss (only between one user and one item). Thus, AECF and CCSR are more likely to be affected by the precision drop of feature values.


\subsection{Similarity loss v.s. Rating reconstruction loss (Q3)}
\label{sec:two_losses}
In this section, we investigate why similarity loss works better for recommendation tasks. To remove the effect of binary loss, we investigate two trained AECF-C and CCSR-C models. We separate test users into two groups: one group that obtained better results in NDCG@10 with CCSR-C, and the other with the worse result. We compute Chi-squares statistics to see which characteristics are effective to separate these groups. The characteristics include average ratings per user, average ratings standard deviation per user, and average number of ratings per user.

\begin{figure}
    \centering
    \includegraphics[width=\linewidth]{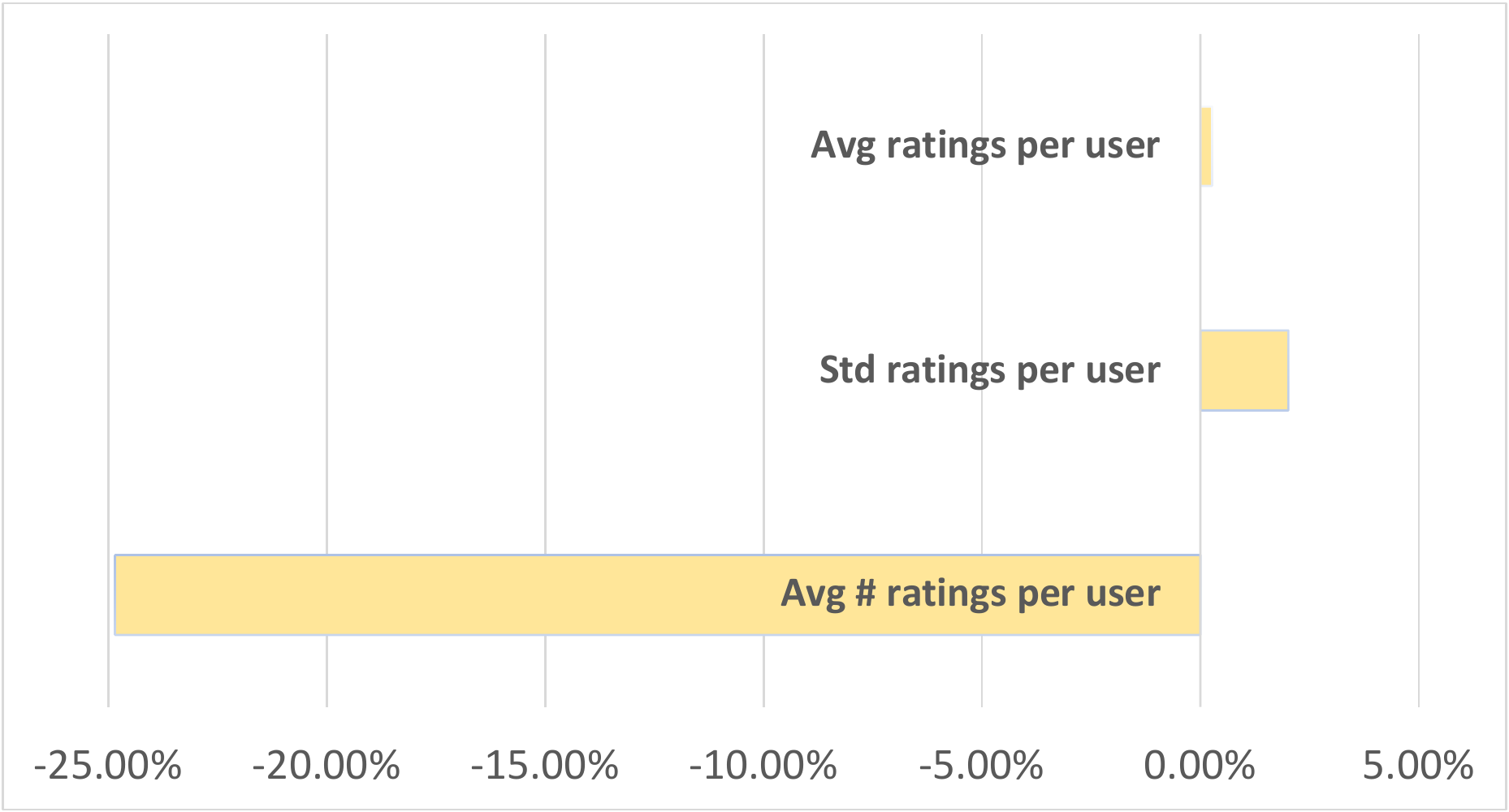}
    \caption{Relative differences of the three characteristics between two user groups with code length 40 on Movielens1M. Users whose performance was improved by CCSR-C model have higher standard deviation of ratings than the ones whose performed was worsen. Users with higher average number of ratings obtained better performance with AECF-C than CCSR-C.}
    \label{fig:betterworse}
\end{figure}

We find some interesting trends as shown in Table~\ref{tab:betterworse}. Most important characteristics to separation is the number of ratings. We check the raw statistics in Figure~\ref{fig:betterworse} and conclude that CCSR is helpful for users who rated less items. AFCF works well for users with many ratings, as it learns the representation to reconstruct original ratings. Considering the sparsity of real-world datasets, we expect similarity loss would be more useful. 

Meanwhile, similarity-based loss works better when the standard deviation in the rating is higher. This makes sense since similarity loss learns the representation using similar and dissimilar pairs. If one user give all items the same ratings, CR-based model will not learn anything useful. 

\begin{table}
\centering
\caption{Chi-squared statistics of user's characteristics in Movielens1M dataset between two user groups: the one whose performance was improved by similarity loss and the other with decreased performance. * indicates the value is statistically significant ($p\leq0.05$)}
\label{tab:betterworse}
\begin{tabular}{c|ccc} 
\toprule
 code length & \# of ratings & 
 avg. ratings  & 
 std. ratings  \\ 
\hline
5 & $47.41^*$ & 0.26 & 0.02 \\
10 & $5.93^*$ & 0.05 & 0.08 \\
20 & 2.37 & 0.33 & 0.53 \\
40 & $8.96^*$ & 0.03 & 0.11 \\
\bottomrule
\end{tabular}
\end{table}

\section{Conclusion}
In this paper, we proposed a new hashing-based RS, Compact Cross-Similarity Recommender. To the best of our knowledge, this is the first work that builds efficient recommender systems from the viewpoint of compact neural cross-modal retrieval. It is encouraging to see the connections between the two research topics. CCSR utilizes autoencoders and MAP similarity to extract features and model interaction between users and items. We demonstrated that CCSR outperforms other MF-based hashing models. We concluded that MAP-based similarity loss is better than MF-based loss in the top-k recommendation task because learning the similarities between user and items are more directly related to the recommendation task compared to learning the ratings. 
From extensive studies of several large-scale datasets, we observed the performance changes on different datasets while using the same model. It suggested us using different models based on data sparsity and data types. 
We also studied different binarization methods and discovered that scaled $tanh$ suffered from the performance drop when its codes were converted to the binary.

\bibliographystyle{ACM-Reference-Format}
\bibliography{sample-base}

\end{document}